\documentclass[twocolumn]{aastex701}

\usepackage{algorithm}
\usepackage{tikz,xcolor,hyperref,colortbl}
\usepackage{graphicx}
\usepackage{amsmath}
\usepackage{bm}
\usepackage{CJKutf8}
\begin{document}

\title{From DESI to \textit{Euclid}: A Generative Bridge to Improve Measurements of Galaxy Structure}

\author[orcid=0000-0002-2339-5581]{Renhao Ye \begin{CJK}{UTF8}{gbsn}(叶人豪)\end{CJK}}
\email{renhaoye@shao.ac.cn}
\affiliation{Shanghai Astronomical Observatory, Chinese Academy of Sciences}
\affiliation{School of Astronomy and Space Science, University of Chinese Academy of Sciences}

\author[orcid=0000-0002-3073-5871]{Shiyin Shen \begin{CJK}{UTF8}{gbsn}(沈世银)\end{CJK}}
\email{ssy@shao.ac.cn}
\affiliation{Shanghai Astronomical Observatory, Chinese Academy of Sciences}
\affiliation{Shanghai Key Lab for Astrophysics, Shanghai Normal University}

\begin{abstract}
Ground-based seeing imprints size-dependent biases on galaxy structural parameters, yet the high-resolution space-based imaging needed to improve these measurements currently covers only a small fraction of the sky. We close this gap with a generative model that produces \textit{Euclid}-like VIS predictions from DESI imaging of Bright Galaxy Survey (BGS) targets. Our predictions remain Fourier-correlated with \textit{Euclid} VIS images down to $0\farcs37$, compared with $1\farcs41$ and $1\farcs00$ for the DESI $r$- and $z$-band inputs, corresponding to improvements by factors of ${\sim}3.8$ and ${\sim}2.7$, respectively. Although this correlation does not extend down to $0\farcs16$, the characteristic \textit{Euclid} VIS PSF FWHM, structural measurements from these predictions already show reduced biases relative to the DESI $r$-band structure measurements: the Petrosian radius bias falls to $+0\farcs072$ (from $-0\farcs845$), independent of galaxy size; the bias in the S\'ersic effective radius ($R_{\rm e}$) drops to $-0\farcs018$ (from $-0\farcs322$); and the S\'ersic-index bias to $+0.093$ (from $+0.262$). We release these predictions over the \textit{Euclid} DR1 footprint as the \textit{Euclid}-like Predictions of BGS (\textbf{E-BGS}), which can be blindly validated once DR1 is public.
\end{abstract}


\section{Introduction}
\label{sec:intro}
Galaxy structure traces the spin, gas cooling, and assembly history of dark matter halos. At the population level, this physics is read off scaling relations such as the mass--size relation \citep{shen_size_2003}, along which star-forming and quiescent galaxies follow separate sequences that evolve with redshift \citep{wel_3dhst+candels_2014,mowla_cosmosdash_2019}, marking out channels of structural growth. Mapping these channels requires accurate structural measurements of the compact and dwarf galaxies at their extremes. The DESI BGS \citep{hahn_desi_2023b} now provides far larger samples of both at low redshift than were previously available.

These BGS galaxies, however, are imaged from the ground, where atmospheric seeing limits resolution: a quarter of BGS galaxies have $r$-band half-light radii\footnote{Half-light radii are taken as \texttt{SHAPE\_R} from the official Tractor catalog.} smaller than the median full width at half maximum (FWHM) of the point-spread function (PSF; $\approx1\farcs3$; \citealt{dey_overview_2019a}), leaving them under-resolved. In this regime, structural parameters acquire size-dependent systematic biases \citep{trujillo_effects_2001}. Space-based imaging directly mitigates this limitation: \textit{Euclid}'s Wide Survey reaches $\mathrm{FWHM}=0\farcs16$ in the Visible Instrument (VIS) band \citep{euclidcollaboration_euclid_2024c, collaboration_euclid_2025m} and will ultimately cover the full BGS footprint, but so far overlaps only ${\sim}63.1\,\mathrm{deg}^2$ of it, rising to the ${\sim}1{,}900\,\mathrm{deg}^2$ of the first \textit{Euclid} Data Release (DR1) in November 2026. The vast majority of BGS galaxies therefore still lack the high-resolution imaging needed to improve structural measurements.

Measuring galaxy structure from under-resolved imaging has long relied on accounting for the PSF at the measurement stage. Analytic prescriptions correct raw S\'ersic parameters for the seeing-convolved profile \citep{trujillo_effects_2001}; two-dimensional forward modeling instead convolves a parametric model with the measured PSF before comparing it with the observations \citep{haussler_gems_2007a, haussler_megamorph_2013, vanderwel_structural_2012a, gao_bulgedisc_2017, wang_psf_2024}; and artificial-degradation experiments calibrate the remaining biases empirically \citep{giavalisco_morphology_1996, yu_redshifting_2023, sazonova_statmorphlsst_2026}. Because forward models explicitly include PSF convolution, the PSF FWHM is not a hard lower limit on measurable galaxy size: effective radii below this scale can remain constrained when the PSF is accurately characterized, the signal-to-noise ratio is sufficient, and the adopted profile adequately describes the galaxy. Euclid forecast studies further show that the accuracy of structural parameters depends on profile realism and the adopted fitting priors \citep{euclidcollaboration_bretonniere_2022, euclidcollaboration_bretonniere_2023}. These limitations motivate our test of whether a learned DESI-to-\textit{Euclid} mapping can improve agreement with observed \textit{Euclid} VIS structural measurements.

Data-driven generative models offer a complementary route: rather than applying a deterministic correction to a single degraded image, they learn the conditional distribution between ground- and space-based observations and generate predictions containing finer-scale structure than the raw ground-based inputs \citep{schawinski_generative_2017, akhaury_deep_2022, zhang_asbridge_2026}. This mapping is intrinsically underdetermined: the predictions can combine object-specific, low-spatial-frequency information from the observed galaxy with morphological correlations learned from the training population, particularly when the input signal-to-noise ratio is low \citep{adam_bayesian_2024, ruan_investigation_2025c}. Individual predicted fine-scale features therefore need not be uniquely specified by the DESI pixels. The appropriate validation target is not to prove feature-by-feature recovery or to rule out a contribution from the learned prior, but to quantify predictive agreement with independent, held-out \textit{Euclid} observations and to state explicitly the limitations of that agreement.

In this work, we train a bridge diffusion model \citep{liu_i2sb_2023} on observed DESI--\textit{Euclid} image pairs from \textit{Euclid} Quick Data Release 1 (Q1) over their $63.1\,\mathrm{deg}^2$ overlap, learning a conditional mapping from DESI to \textit{Euclid}-like predictions. We evaluate it against two criteria: whether the predictions exhibit correlation with \textit{Euclid} VIS observations on smaller angular scales than the raw DESI inputs, and whether the structural parameters measured from them agree with the corresponding \textit{Euclid} measurements. We then release the resulting predictions over the \textit{Euclid} DR1 footprint as E-BGS, an advance prediction testable once DR1 is public. The paper is organized as follows: Sections~\ref{sec:data} and~\ref{sec:method} describe the data, sample construction, and the method; Section~\ref{sec:results} presents the validation against these two criteria in turn; Section~\ref{sec:conclusion} concludes.

\section{Data}
\label{sec:data}

\subsection{DESI}
\label{sec:desi}
\begin{figure}
    \centering
    \includegraphics[width=1\columnwidth]{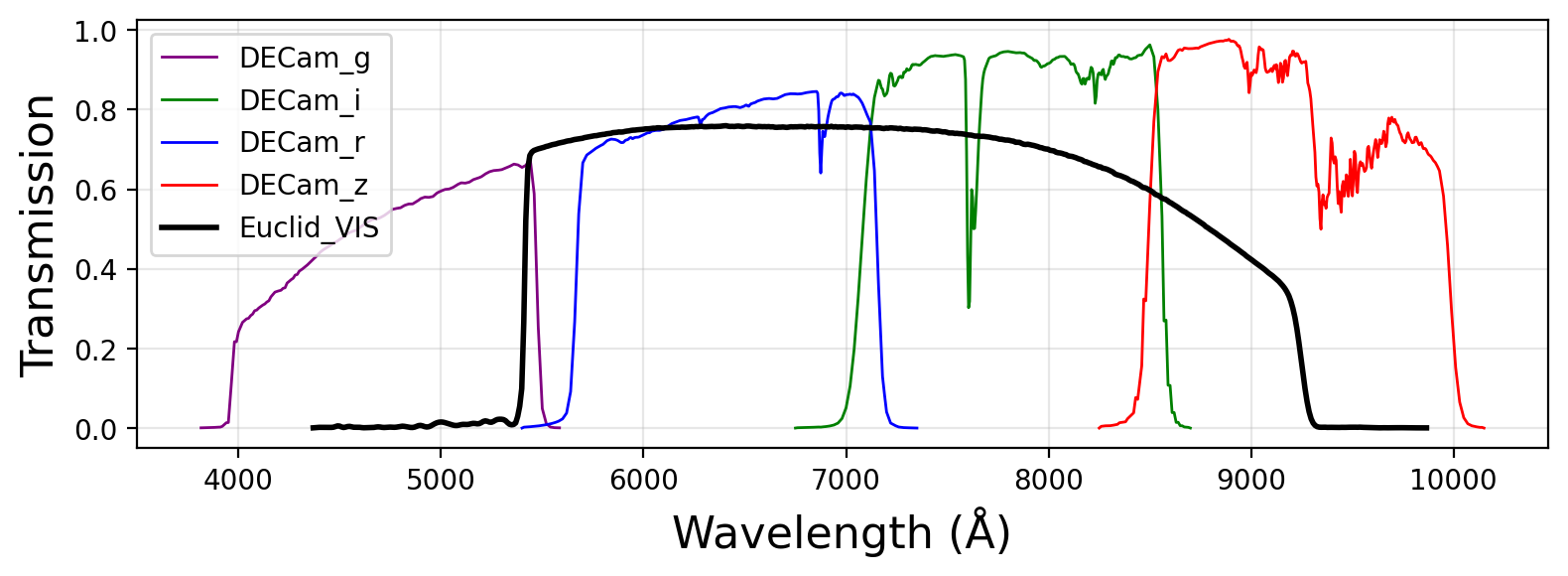}
    \caption{Transmission curves for the DECam $g$, $r$, $i$, and $z$ bands and the \textit{Euclid} VIS channel (${\sim}550$--$900\,\mathrm{nm}$). The DECam curves illustrate the wavelength coverage in the southern footprint; BASS/MzLS use their corresponding northern $grz$ filters.}
    \label{fig:filter}
\end{figure}

The DESI Legacy Imaging Surveys \citep[LIS;][]{dey_overview_2019a} comprise DECaLS (DR10, southern footprint) and BASS/MzLS \citep{zou_project_2017,silva_mayall_2016} (DR9, northern footprint), which together provide $grz$ imaging at $0\farcs262\,\mathrm{pixel}^{-1}$; DR10 additionally includes $i$-band imaging in the southern footprint. We use only the $r$ and $z$ bands (Fig.~\ref{fig:filter}): the $g$ band lies almost entirely below the \textit{Euclid} VIS bandpass, and the $i$ band is dropped for footprint uniformity because it is unavailable over BASS/MzLS.

We draw our sample from the DESI BGS targets \citep{hahn_desi_2023b}, which split into BGS Bright ($r < 19.5$, flux-limited) and BGS Faint ($19.5 < r < 20.175$, with a fiber-magnitude--color cut for redshift efficiency).

\subsection{\textit{Euclid}}
\label{sec:euclid_q1}
The \textit{Euclid} mission \citep{laureijs_euclid_2011,euclidcollaboration_euclid_2024c} performs a wide-area survey with the VIS imager \citep{collaboration_euclid_2025m} and the Near-Infrared Spectrometer and Photometer (NISP). Q1 \citep{collaboration_euclid_2025e} provides $63.1\,\mathrm{deg}^2$ over three Euclid Deep Fields at nominal Wide Survey depth ($I_{\rm E}=26.2\,\mathrm{mag}$, $5\sigma$, point source). We use the VIS broadband ($I_{\rm E}$, ${\sim}550$--$900\,\mathrm{nm}$) on a $0\farcs1\,\mathrm{pixel}^{-1}$ grid, a single channel spanning the DESI $r$, $i$, and $z$ bands (Fig.~\ref{fig:filter}). These space-based VIS images provide the \textit{Euclid} observational reference against which we evaluate the predictions and their measured structural parameters. They are not treated as the intrinsic, PSF-free galaxy truth: the measured morphology remains dependent on the VIS bandpass and is affected by the \textit{Euclid} PSF, measurement noise, and MERge (MER) data product \citep{euclid_mer_2026}.

\subsection{Sample Construction}
\label{sec:preproc}
Each BGS target in the Q1 footprint is matched to a VIS-detected, non-spurious, non-stellar source of MER  product (full flag list in Appendix~\ref{app:selection}), giving paired DESI--\textit{Euclid} imaging for every galaxy. We partition the pairs by sky region rather than by random draw. The test set is the region where Q1 and DESI DR1 overlap; the remaining, spatially disjoint pairs form the training and validation sets.

For each galaxy, DESI $r$ and $z$ bricks are reprojected onto the \textit{Euclid} VIS pixel grid ($0\farcs1\,\mathrm{pixel}^{-1}$) and $128\times128$-pixel ($12\farcs8$) cutouts are extracted from both surveys with \textsc{Cutout2D} \citep{collaboration_astropy_2022}, all centered on the common \textit{Euclid} MER position (reprojection and multi-brick handling in Appendix~\ref{app:preproc}).

This yields $82{,}830$ photometrically selected pairs for training and validation, split $80{:}20$ into $66{,}359$ and $16{,}471$; the test region contributes a further $4{,}165$ pairs.

\begin{figure*}
    \centering
    \includegraphics[width=2\columnwidth]{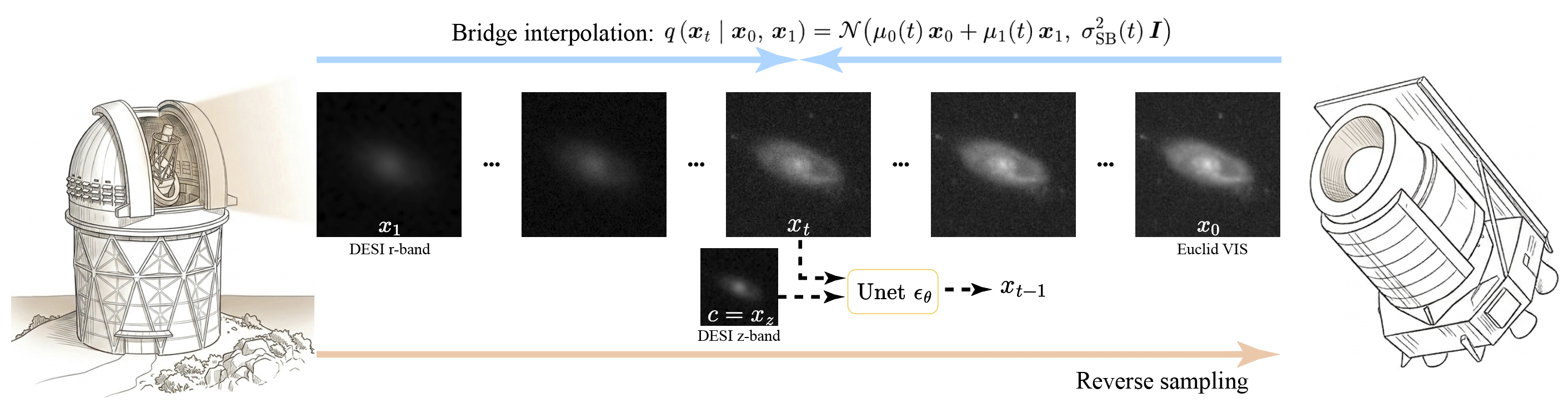}
    \caption{The I$^2$SB framework used to generate \textit{Euclid}-like VIS predictions from DESI. The image sequence runs between two fixed endpoints, the DESI $r$-band source $\bm{x}_1$ (left) and the \textit{Euclid} VIS target $\bm{x}_0$ (right), through intermediate states $\bm{x}_t$. The blue arrow marks the forward bridge interpolation (Eq.~\ref{eq:sb_forward}); the orange arrow marks reverse sampling, in which the U-Net $\epsilon_\theta$, conditioned on the DESI $z$-band ($\bm{c}=\bm{x}_z$), predicts an estimate of the \textit{Euclid} endpoint from $\bm{x}_t$, and the posterior update produces the next reverse state. Intermediate frames are model states, not observations.}
    \label{fig:overview}
\end{figure*}
\section{Methods}
\label{sec:method}

\subsection{A Bridge Between Two Real Observations}

We model the DESI-to-\textit{Euclid} mapping with the Image-to-Image Schr\"odinger Bridge \citep[I$^2$SB;][]{liu_i2sb_2023}, which connects two fixed image endpoints by a continuous path (Figure~\ref{fig:overview}). We define this path in the forward direction (the \emph{forward process}) as the gradual blurring of the \textit{Euclid} image into its DESI counterpart, and train a network to reverse it, producing a prediction of the \textit{Euclid} image conditioned on the DESI input.

This process is a stochastic bridge between $\bm{x}_0$ (\textit{Euclid}) and $\bm{x}_1$ (DESI). For $t=0,\ldots,T-1$, each intermediate state $\bm{x}_t$ is drawn from a Gaussian distribution whose mean is a weighted combination of the two endpoint images and whose variance $\sigma_{\rm SB}^2(t)$ sets its stochastic spread:
\begin{equation}
    q\!\left(\bm{x}_t \mid \bm{x}_0,\, \bm{x}_1\right)
    = \mathcal{N}\!\left(
        \mu_0(t)\,\bm{x}_0 + \mu_1(t)\,\bm{x}_1,\;
        \sigma^2_{\mathrm{SB}}(t)\,\bm{I}
      \right),
    \label{eq:sb_forward}
\end{equation}
where $\bm{I}$ is the identity matrix and the coefficients are derived from the forward and backward cumulative standard deviations, summed over the schedule steps $s$,
$\sigma_{\mathrm{fwd}}(t) = \sqrt{\sum_{s \leq t} \beta_s}$ and
$\sigma_{\mathrm{bwd}}(t) = \sqrt{\sum_{s > t} \beta_s}$:
\begin{align}
    \mu_0(t) &= \frac{\sigma_{\mathrm{bwd}}^2(t)}{\sigma_{\mathrm{fwd}}^2(t) + \sigma_{\mathrm{bwd}}^2(t)},
    \qquad
    \mu_1(t) = \frac{\sigma_{\mathrm{fwd}}^2(t)}{\sigma_{\mathrm{fwd}}^2(t) + \sigma_{\mathrm{bwd}}^2(t)},
    \nonumber \\
    \sigma_{\mathrm{SB}}^2(t) &= \frac{\sigma_{\mathrm{fwd}}^2(t)\,\sigma_{\mathrm{bwd}}^2(t)}
                                      {\sigma_{\mathrm{fwd}}^2(t) + \sigma_{\mathrm{bwd}}^2(t)}.
    \label{eq:sb_coefs}
\end{align}
The weights shift the mean from a state close to $\bm{x}_0$ to one close to $\bm{x}_1$ as $t$ increases, while the variance is smallest near the two ends and peaks near the midpoint. Because the discrete schedule has small but nonzero endpoint values, the first and last indexed bridge distributions approach rather than exactly equal the endpoints. The stochastic bridge provides a family of intermediate states whose reverse supports the one-to-many nature of predicting fine-scale structure from a blurred input. The schedule $\{\beta_s\}$ setting these terms is given in Appendix~\ref{app:implementation}.

\subsection{Network Input and Training Objective}

A U-Net with trainable parameters $\theta$ learns the reverse path one step at a time. Given the current state $\bm{x}_t$, the step $t$, and the conditioning image, it predicts the normalized residual $\bm{\epsilon}=(\bm{x}_t-\bm{x}_0)/\sigma_{\mathrm{fwd}}(t)$, from which the \textit{Euclid} endpoint is estimated as $\widehat{\bm{x}}_0=\bm{x}_t-\sigma_{\mathrm{fwd}}(t)\bm{\epsilon}_\theta(\bm{x}_t,\bm{c},t)$. The DESI $r$-band image defines the bridge source $\bm{x}_1$, while the DESI $z$-band image enters as a second channel ($\bm{c}=\mathrm{DESI}_z$) concatenated with $\bm{x}_t$. The per-band arcsinh stretch detailed in Appendix~\ref{app:implementation} places these inputs on fixed numerical ranges. The network is trained under a masked mean squared error loss:
\begin{equation}
    \mathcal{L}(\theta)
    = \mathbb{E}\!\left[
        {\frac{
        \sum_i M_i
        {\color{black}\left|
            \epsilon_\theta\!\left(\bm{x}_t,\, \bm{c},\, t\right){_i}
            - \left(\frac{\bm{x}_t - \bm{x}_0}{\sigma_{\mathrm{fwd}}(t)}\right){_i}
        \right|^2}
        }{\max\!\left(\sum_i M_i,1\right)}}
      \right],
    \label{eq:loss}
\end{equation}
where $i$ indexes pixels and $M_i$ is the binary validity mask described in Appendix~\ref{app:implementation}. At inference, reverse sampling is initialized exactly at the DESI image $\bm{x}_1$. At each selected reverse step, the network estimates $\bm{x}_0$ and the bridge posterior produces the next state. The network architecture is specified in Appendix~\ref{app:implementation}.

\begin{figure*}
    \centering
    \includegraphics[width=2.1\columnwidth]{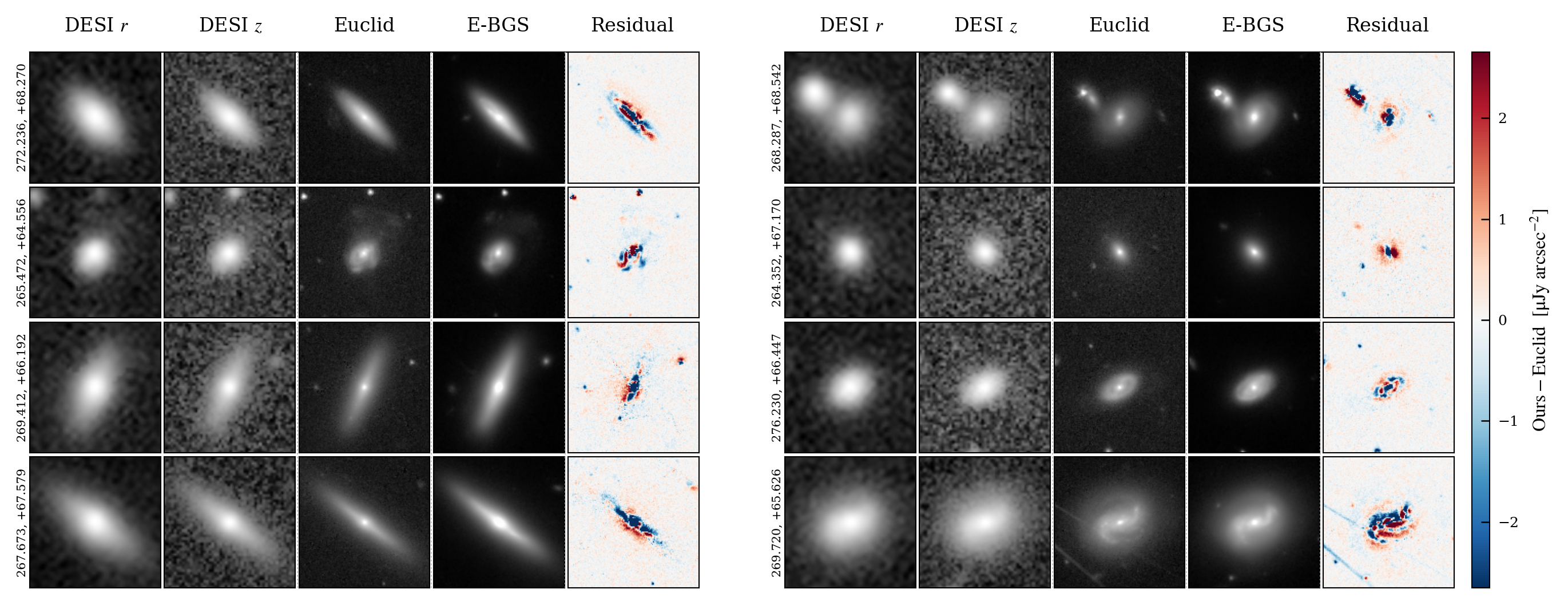}
    \caption{Predictions for randomly selected test-set galaxies (labeled by R.A., Dec.). Columns: DESI $r$-band, DESI $z$-band, \textit{Euclid} Q1, prediction, and residual (prediction~$-$~\textit{Euclid}).}
    \label{fig:rec}
\end{figure*}

\begin{figure}
    \centering
    \includegraphics[width=\columnwidth]{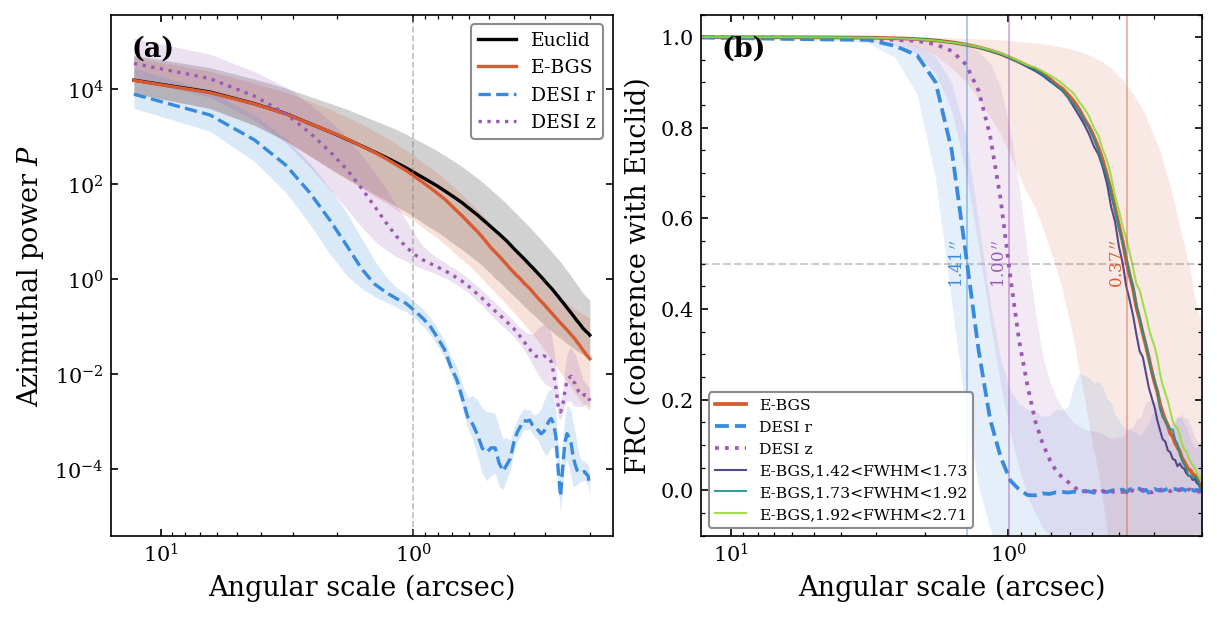}
    \caption{Fourier-domain comparison with the held-out \textit{Euclid} VIS observations over the test set (median; shaded 16th--84th-percentile intervals). \emph{(a)} Azimuthally averaged power spectrum $P$ versus angular scale for \textit{Euclid} VIS, E-BGS, and the DESI $r$ and $z$ bands. \emph{(b)} FRC with the held-out \textit{Euclid} VIS observations for E-BGS and the DESI bands. The horizontal dashed line marks $\mathrm{FRC}=0.5$, and the vertical dashed lines mark the corresponding crossing scale $q_{0.5}$ for each curve. Thin curves show the E-BGS FRC in three bins of DESI $r$-band PSF FWHM.}
    \label{fig:freq}
\end{figure}

\begin{figure*}
    \centering
    \includegraphics[width=2\columnwidth]{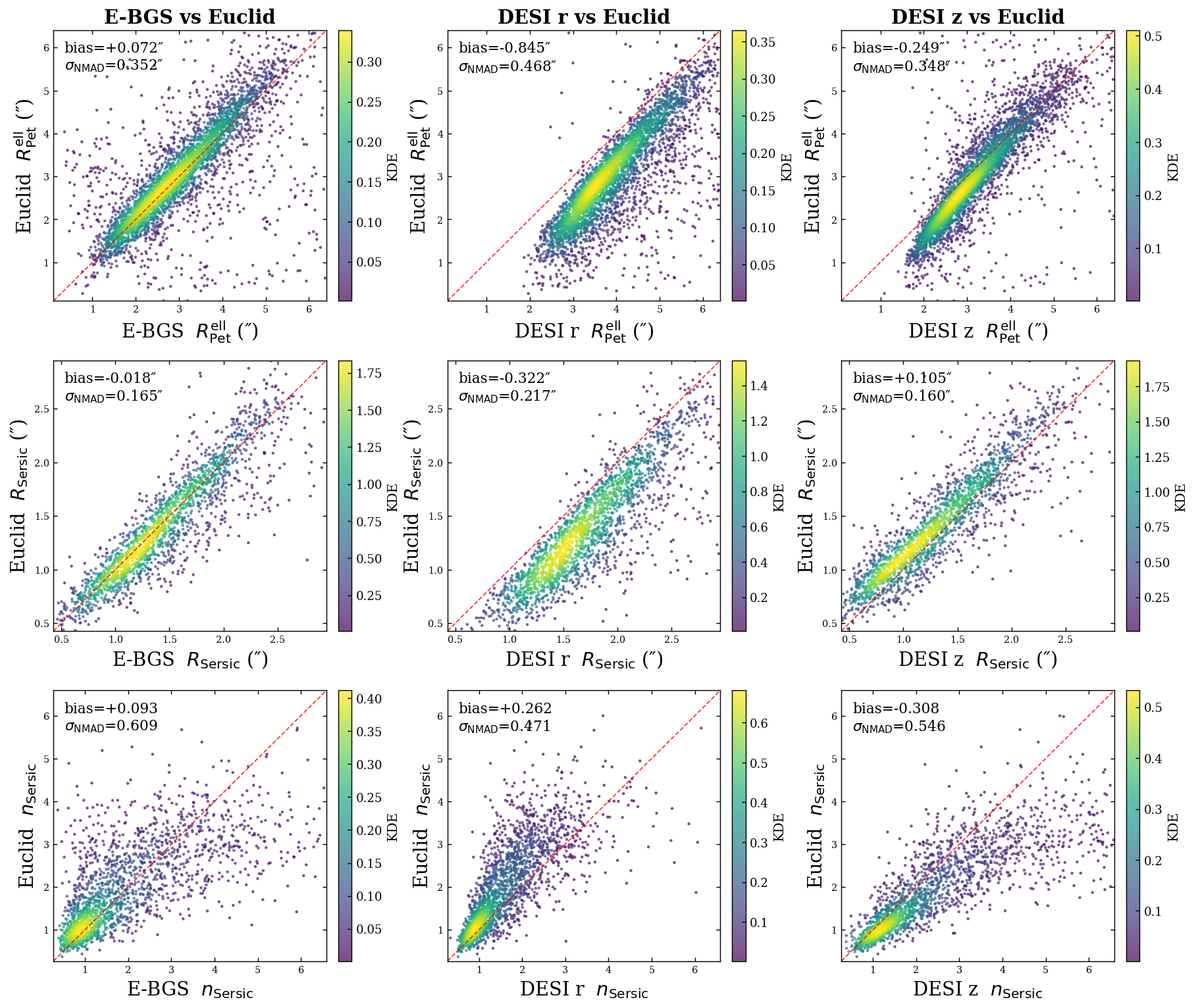}
    \caption{Structural parameters relative to \textit{Euclid}, one per row: elliptical Petrosian radius $R_{\rm Pet}$ (top), S\'ersic effective radius $R_{\rm e}$ (middle), and S\'ersic index $n$ (bottom). Columns compare E-BGS (left), DESI $r$-band (center), and DESI $z$-band (right) with the corresponding \textit{Euclid} measurements; points are color-coded by the kernel density estimate (KDE), and the dashed red line is the one-to-one relation\textcolor{red}.
    Inset: median bias and normalized median absolute deviation (NMAD) scatter $\sigma_{\rm NMAD}$ of the residual $\Delta\equiv\textit{Euclid}-\mathrm{method}$.}
    \label{fig:size_sys}
\end{figure*}

\section{Results}
\label{sec:results}

\subsection{Example Predictions}
\label{sec:rec}

Figure~\ref{fig:rec} shows randomly selected test-set galaxies. Compared with the DESI $r$- and $z$-band images, the predictions appear visually closer to the corresponding \textit{Euclid} VIS observations and show finer morphological features resembling those in the \textit{Euclid} images, including spiral arms, nearby companion galaxies and disk substructure. Because the DESI images contain little direct information about these fine-scale features, the conditional mapping is underdetermined: the model must infer such details by combining object-specific, low-spatial-frequency information from the DESI observations with a data-driven morphological prior learned from the \textit{Euclid} images in the training set. As a result, the predicted small-scale features do not always appear at the same locations as their counterparts in the \textit{Euclid} observations, producing strong pixel-level residuals.

The predictions are generally smoother than the \textit{Euclid} observations but sharper than the DESI images, as illustrated by the bulge of the edge-on galaxy in the last panel of the first row in Figure~\ref{fig:rec}. Their visual agreement with \textit{Euclid} is therefore incomplete at fine spatial scales. In the following subsection, we quantify this agreement using the Fourier-domain correlation between the predictions and the held-out \textit{Euclid} observations.

Separately, a random selection of the released E-BGS predictions over the \textit{Euclid} DR1 footprint, where no corresponding public \textit{Euclid} imaging is currently available, is shown in Figure~\ref{fig:ebgs_release}. Unlike the paired test cutouts, which are centered on \textit{Euclid} MER positions, the released predictions use DESI $r$-band Tractor positions because \textit{Euclid}-derived positions are unavailable at deployment.

\subsection{Structural Fidelity}
\label{sec:fidelity}
Predicting fine-scale structure from seeing-limited ground-based images is an ill-posed inverse problem with multiple possible solutions. A direct pixel-by-pixel comparison between our predictions and the corresponding \textit{Euclid} images does not by itself fully characterize their structural agreement. We therefore also assess this agreement in the frequency domain using two complementary diagnostics (Figure~\ref{fig:freq}): the azimuthally averaged power spectrum to quantify Fourier power at each spatial scale, and the Fourier Ring Correlation \citep[FRC;][]{vanheel_fourier_2005} to quantify the correlation of the complex Fourier coefficients with \textit{Euclid}.

We compute both diagnostics for all $4{,}165$ test galaxies, as no non-finite pixel values are present in any of the E-BGS, DESI $r$-band, DESI $z$-band, or \textit{Euclid} images. Before Fourier transformation, we subtract the median background from each cutout and apply neither additional pixel masking nor flux normalization. We then apply the same two-dimensional Hann window to each cutout to suppress spectral leakage from the image boundaries. After computing the two-dimensional Fourier transform, we group the coefficients into radial rings one pixel wide and retain spatial frequencies up to $5\,\mathrm{arcsec}^{-1}$. This corresponds to an angular scale of $0\farcs2$, twice the $0\farcs1$ pixel size and therefore the finest scale that can be represented on this pixel grid without aliasing. The azimuthally averaged power spectrum is
\begin{equation}
P(k)=\frac{1}{N_k}\sum_{\mathcal{R}_k}|F|^2,
\end{equation}
where $\mathcal{R}_k$ denotes the ring at spatial frequency $k$ and $N_k$ is the number of Fourier coefficients in that ring.

We measure the correlation with \textit{Euclid} within the same rings using
\begin{equation}
\mathrm{FRC}(k)=
\frac{
\sum_{\mathcal{R}_k}
\mathrm{Re}\!\left(F_{\rm method}F_{\rm Euclid}^{*}\right)
}{
\sqrt{
\sum_{\mathcal{R}_k}|F_{\rm method}|^2
\sum_{\mathcal{R}_k}|F_{\rm Euclid}|^2
}
},
\end{equation}
where $F_{\rm method}$ and $F_{\rm Euclid}$ are the Fourier transforms of the method and \textit{Euclid} images, respectively. The normalization removes sensitivity to an overall multiplicative amplitude scaling within a ring, so the FRC evaluates the correlation of the complex Fourier coefficients rather than their absolute power. We compute the diagnostics object by object and report their sample medians and 16th--84th percentiles at each spatial frequency. With the $0\farcs1\,\mathrm{pixel}^{-1}$ sampling, the angular scale corresponding to spatial frequency $k$ is $q=1/k$. We define $q_{0.5}$ as the angular scale at which the median FRC first falls below 0.5 as spatial frequency increases. Excluding the zero-frequency point, we locate this crossing by linear interpolation between adjacent rings.

Relative to \textit{Euclid}, the DESI $r$ and $z$ images contain up to three orders of magnitude less Fourier power at angular scales below approximately $1''$ (Figure~\ref{fig:freq}a). E-BGS contains substantially more high-spatial-frequency power than the DESI inputs and lies closer to the \textit{Euclid} spectrum, although its power remains approximately $0.5\,\mathrm{dex}$ lower. This offset is measured as the median difference in $\log_{10}P$ between the E-BGS and \textit{Euclid} curves over $0\farcs2$--$1''$, where the DESI inputs are most strongly degraded. Because comparable Fourier power does not establish whether structures occur at the same locations, we use the FRC to assess their correlation with \textit{Euclid}.

The median FRC between E-BGS and \textit{Euclid} crosses 0.5 at $q_{0.5}=0\farcs37$, compared with $1\farcs41$ for DESI $r$ and $1\farcs00$ for DESI $z$ (Figure~\ref{fig:freq}b). The E-BGS crossing therefore occurs at angular scales approximately 3.8 and 2.7 times smaller than those of the DESI $r$- and $z$-band inputs, respectively. Because the three curves remain nearly parallel across the plotted range, these relative factors are largely insensitive to the particular FRC threshold used to define the crossing. Although the finer crossing scale of DESI $z$ is consistent with its generally narrower PSF, $q_{0.5}$ is an empirical FRC crossing scale rather than a measurement of the PSF FWHM. The higher E-BGS FRC at small angular scales demonstrates stronger predictive Fourier correlation with \textit{Euclid} than is present in the raw DESI inputs, but it does not establish feature-by-feature recovery or exclude contributions from the learned conditional prior. Across the three DESI $r$-band seeing bins, the E-BGS crossing scale varies by approximately $0\farcs04$.

\subsection{Structural-Parameter Agreement}
\label{sec:morph}
We evaluate three structural parameters: the elliptical Petrosian radius \citep[$R_{\rm Pet}$;][]{petrosian_surface_1976, graham_concise_2005}, the S\'ersic effective radius ($R_{\rm e}$), and the S\'ersic index ($n$). For each parameter, we define the residual as $\Delta\equiv\textit{Euclid}-\mathrm{method}$, the bias as the sample $\mathrm{median}(\Delta)$, and the scatter as the normalized median absolute deviation (NMAD), $\sigma_{\rm NMAD}=1.4826\,\mathrm{median}(|\Delta-\mathrm{median}(\Delta)|)$.

We measure size first with the Petrosian radius (top row of Figure~\ref{fig:size_sys}). It is non-parametric, assuming no form for the light profile, but it cannot account for the way the PSF spreads the galaxy light outward, so the Petrosian radius overestimates the size \citep{trujillo_effects_2001, geda_petrofit_2022}. Because a centered profile can extend only $6\farcs4$ before reaching the edge of a $12\farcs8$ cutout, we exclude each Petrosian comparison for which either the \textit{Euclid} or comparison radius exceeds $6\farcs4$. This excludes 201 of 4,165 galaxies for E-BGS, 403 for DESI $r$, and 175 for DESI $z$. After this exclusion, the DESI $r$ and $z$ bands show median residuals of $\Delta R_{\rm Pet}=-0\farcs845$ and $-0\farcs249$, respectively. In DESI $r$, the overestimate is larger for smaller galaxies. E-BGS reduces the sample-wide median offset to $\Delta R_{\rm Pet}=+0\farcs072$, and the size-dependent trend is substantially weaker in the tested sample.

The S\'ersic effective radius $R_{\rm e}$ and index $n$ are fitted with \texttt{statmorph} on a clean subset defined by a S\'ersic-fit $\chi^2_\nu<3$ cut on the \textit{Euclid} reference. This cut retains 2,553 of the 4,165 test galaxies (61.30\%) and likely underrepresents irregular, merging, and poorly described single-S\'ersic systems. Because it requires the \textit{Euclid} reference fit, the same selection cannot be applied in regions without \textit{Euclid} data. The DESI and \textit{Euclid} images are fitted with their measured PSFs, while E-BGS is fitted using randomly drawn \textit{Euclid} Q1 PSFs; the implications and sensitivity of this approximation are discussed later. For $R_{\rm e}$ (middle row of Figure~\ref{fig:size_sys}), the DESI measurements track the \textit{Euclid} measurements along the one-to-one line, offset from it only by a constant; E-BGS yields a small sample-wide median residual of $\Delta R_{\rm e}=-0\farcs018$, compared with $-0\farcs322$ for DESI $r$ and $+0\farcs105$ for DESI $z$. For $n$ (bottom row), the DESI offset is value-dependent rather than a constant shift, larger at high $n$, where DESI $r$ underestimates and $z$ overestimates. Their sample-wide median residuals are $\Delta n=+0.262$ and $-0.308$, respectively. E-BGS reduces the median residual to $\Delta n=+0.093$ and weakens the value dependence within the selected test sample.

E-BGS predictions are not direct observations and therefore do not have directly measured physical PSFs. Assigning each prediction a randomly drawn \textit{Euclid} Q1 PSF is an operational approximation for the PSF-convolved S\'ersic fitting used in this work. Accordingly, the reported reductions in systematic offsets are conditional on this PSF-assignment procedure. Repeating the fit with 10 independently drawn PSFs for each of the 2,553 galaxies in the selected test sample yields a median within-galaxy NMAD of $0\farcs00070$ for $R_{\rm e}$ and 0.00184 for $n$, indicating that the particular PSF drawn contributes little additional scatter within the tested PSF distribution.

Although the structural measurements in the DESI $r$ and $z$ bands show clear systematic differences relative to \textit{Euclid} VIS, these differences cannot be attributed entirely to the lower resolution or measurement limitations of the DESI images. The comparison is not strictly bandpass matched: the broad \textit{Euclid} VIS band approximately spans the wavelength range covered by the DESI $r$, $i$, and $z$ bands, whereas the DESI structural parameters are measured separately in $r$ and $z$. The systematic differences may therefore arise partly from color gradients, dust attenuation, and differences between the stellar populations of bulges and disks. Under these qualifications, the reductions in median residuals reported above should be interpreted as improved sample-wide agreement between E-BGS and the observed \textit{Euclid} VIS structural measurements, rather than as recovery of an intrinsic, bandpass-independent galaxy morphology. This improvement in agreement, as measured by the median residual, does not generally extend to the object-to-object scatter. After the Petrosian boundary exclusion, the E-BGS NMAD is $0\farcs352$ for $R_{\rm Pet}$, compared with $0\farcs468$ for DESI $r$ and $0\farcs348$ for DESI $z$; its NMAD is $0\farcs165$ for $R_{\rm e}$, compared with $0\farcs160$ for DESI $z$; and 0.609 for $n$, compared with 0.471 for DESI $r$ and 0.546 for DESI $z$. E-BGS therefore does not yield the lowest residual scatter for any of the three parameters.

\section{Conclusion}
\label{sec:conclusion}

We trained a bridge diffusion model on real DESI--\textit{Euclid} Q1 image pairs to generate \textit{Euclid}-like VIS predictions from DESI imaging. On the spatially disjoint test set, E-BGS remains Fourier-correlated with the held-out \textit{Euclid} VIS observations down to $0\farcs37$ at $\mathrm{FRC}=0.5$, compared with $1\farcs41$ and $1\farcs00$ for the DESI $r$- and $z$-band inputs, corresponding to improvements by factors of ${\sim}3.8$ and ${\sim}2.7$ in the FRC crossing scale, respectively. Because the mapping from seeing-limited DESI images to \textit{Euclid}-like predictions is underdetermined, this result does not imply that the fine-scale morphology of an individual prediction is uniquely constrained by the DESI input; it may also reflect morphological correlations learned from the paired training sample.

Under the adopted PSF-assignment, fitting, and sample-selection procedures, the sample-wide median residuals for $(R_{\rm Pet}, R_{\rm e}, n)$ are $(+0\farcs072,-0\farcs018,+0.093)$ for E-BGS, compared with $(-0\farcs845,-0\farcs322,+0.262)$ for DESI $r$ and $(-0\farcs249,+0\farcs105,-0.308)$ for DESI $z$. E-BGS therefore provides closer sample-wide median agreement with the observed \textit{Euclid} VIS measurements for all three parameters, although it does not yield the lowest NMAD for any of them. These results should not be interpreted as recovery of intrinsic, bandpass-independent morphology and do not exclude conditional biases across galaxy subpopulations or biases in population scaling relations. We release E-BGS predictions for BGS targets in the \textit{Euclid} DR1 footprint for blind validation when DR1 becomes public.

\section*{Acknowledgements}
This work was supported by the National Key R\&D Program of China  No.\ 2025YFF0510603
, the National Natural Science Foundation of China (No. 12141302) and the China Manned Space Project (CMS-CSST-2025-A07).

This work has made use of \textit{Euclid} Quick Data Release 1 (Q1) data from the \textit{Euclid} mission of ESA, 2025, \url{https://doi.org/10.57780/esa-2853f3b}, and of ESA Datalabs (\url{datalabs.esa.int}).

The Legacy Surveys (DECaLS, BASS, MzLS) used data obtained at the Blanco telescope (CTIO/NOIRLab), the Bok telescope (Steward Observatory), and the Mayall telescope (KPNO/NOIRLab). Pipeline processing was supported by NOIRLab and LBNL. This project used data obtained with DECam, constructed by the DES collaboration; funding for DES was provided by the U.S.\ Department of Energy, NSF, and numerous international agencies (full list at \url{https://www.darkenergysurvey.org}). BASS is a key project of the Telescope Access Program (TAP), funded by NAOC and the Chinese Academy of Sciences. The Legacy Surveys team thanks the Tohono O'odham Nation for access to Iolkam Du'ag (Kitt Peak). NOIRLab is operated by AURA under a cooperative agreement with NSF; LBNL is managed by the Regents of the University of California under contract to the U.S.\ Department of Energy. The Legacy Surveys imaging of the DESI footprint is supported by DOE Contract No.\ DE-AC02-05CH1123 and NSF Contract No.\ AST-0950945.

\section*{Data Availability}
The version of the training and inference code used in this work is archived on Zenodo \citep{ye_ebgs_2026}; the living code is available at \url{https://github.com/Rh-YE/EBGS}. The E-BGS predictions over the \textit{Euclid} DR1 footprint are available at \url{https://doi.org/10.5281/zenodo.21032414}.

\bibliography{sample701}{}
\bibliographystyle{aasjournalv7}


\appendix

\section{Implementation Details}
\label{app:implementation}

\subsection{Noise Schedule}

The reported model uses a symmetric square-root-linear noise schedule \citep{liu_i2sb_2023} with $T=500$ steps. We first construct an auxiliary sequence between $\beta_{\rm start}=10^{-10}$ and $\widetilde{\beta}_{\rm end}=\beta_{\rm scale}/T=3\times10^{-4}$, where $\beta_{\rm scale}=0.15$ is stored under the legacy configuration name \texttt{beta\_max}:
\begin{equation}
\widetilde{\beta}_j =
\left[
\sqrt{\beta_{\rm start}}+
\frac{j}{T-1}
\left(
\sqrt{\widetilde{\beta}_{\rm end}}-
\sqrt{\beta_{\rm start}}
\right)
\right]^2,
\qquad j=0,\ldots,T-1.
\end{equation}
The final schedule retains the first half of this sequence and reflects it about the midpoint:
\begin{equation}
\beta_t=\widetilde{\beta}_{\min(t,T-1-t)},
\qquad t=0,\ldots,T-1.
\end{equation}
For the adopted parameters, this construction yields a symmetric schedule with two equal central maxima of $7.4786\times10^{-5}$ and $\sum_t\beta_t=0.0125038$. This schedule was used during training and for all inference runs analyzed in this work.

At inference, we select $K=10$ uniformly spaced reverse steps. For each galaxy, three sampling trajectories share the first $70\%$ of the reverse steps and then evolve independently; the released prediction is their mean.

\subsection{Flux Stretching}

Native flux spans several orders of magnitude, which makes raw images inefficient to train on, so before training we map each band into $[-1,1]$ with a per-band arcsinh stretch and linear rescaling:
\begin{equation}
    f_b(x) = 2\,\frac{\mathrm{arcsinh}\!\left(\tfrac{\mathrm{clip}(x,\,L_b,\,U_b)-m_b}{\sigma_b}\right) - v^{\min}_b}{v^{\max}_b - v^{\min}_b} - 1,
    \label{eq:arcsinh}
\end{equation}
where $m_b$ is the median sky background, $\sigma_b$ is a dataset-level stretch scale derived from background fluctuations, and $(L_b,U_b)$ are band-specific clipping bounds. The corresponding rescaling limits are $v_b^{\min}=\mathrm{arcsinh}[(L_b-m_b)/\sigma_b]$ and $v_b^{\max}=\mathrm{arcsinh}[(U_b-m_b)/\sigma_b]$. All parameters were determined from the training-footprint statistics and held fixed during validation and inference (Table~\ref{tab:stretch}). The inverse transform returns each E-BGS prediction to the native linear flux units of the \textit{Euclid} VIS images.

\subsection{Training Loss}

The masked loss (Equation~\ref{eq:loss}) uses the logical intersection of per-pixel validity checks across the DESI and \textit{Euclid} images and their RMS maps. All three stored DESI image bands must lie within $[-0.05,10]$, the \textit{Euclid} image within $[-0.05,20]$, the DESI RMS values within $[10^{-10},1]$, and the \textit{Euclid} RMS values within $[0,1]$; non-finite values fail these checks. The resulting mask is binary and is not an inverse-variance weighting.

\subsection{Network and Optimization}

The U-Net uses base channel width 64, multipliers $(1,2,4,4)$, 2 residual blocks per resolution, and spatial self-attention at relative resolutions $1/16$ and $1/8$. We optimize with AdamW \citep{loshchilov_decoupled_2017} at learning rate $10^{-4}$, batch size 256, with a 1000-step linear warm-up followed by linear decay to 10\% over $10^5$ steps, and exponential moving average (EMA) decay 0.999. We apply no data augmentation: each pair is center-cropped to $128\times128$ with no rotation or flipping. Key hyperparameters are listed in Table~\ref{tab:hparams}.

\begin{table}
\centering
\caption{Per-band arcsinh stretch parameters (Eq.~\ref{eq:arcsinh}). All values in native flux units.}
\label{tab:stretch}
\begin{tabular}{lcccc}
\hline
Band & $L_b$ & $U_b$ & $m_b$ & $\sigma_b$ \\
\hline
\textit{Euclid} VIS  & $-1.543\times10^{-2}$ & $2.093$ & $3.160\times10^{-4}$ & $1.790\times10^{-2}$ \\
DESI $r$             & $-6.752\times10^{-3}$ & $10.000$ & $2.249\times10^{-4}$ & $1.180\times10^{-1}$ \\
DESI $z$             & $-2.320\times10^{-2}$ & $10.000$ & $5.654\times10^{-4}$ & $1.374\times10^{-1}$ \\
\hline
\end{tabular}
\end{table}

\begin{table}
\centering
\caption{Model and training configuration.}
\label{tab:hparams}
\begin{tabular}{ll@{\hspace{2em}}ll}
\hline
Parameter & Value & Parameter & Value \\
\hline
\multicolumn{2}{l}{\textit{Schr\"odinger Bridge}} & \multicolumn{2}{l}{\textit{Optimization}} \\
$T$                          & 500              & Optimizer          & AdamW \\
Legacy schedule scale (\texttt{beta\_max}) & 0.15 & Learning rate      & $10^{-4}$ \\
Function evaluations         & 10               & Batch size         & 256 \\[2pt]
\multicolumn{2}{l}{\textit{U-Net}}        & Warm-up            & 1000 steps \\
Base channels                & 64               & Decay              & $10^5$ steps to $10\%$ \\
Channel multipliers          & $(1,2,4,4)$      & EMA decay          & $0.999$ \\
Residual blocks / resolution & 2                & Weight decay       & $10^{-5}$ \\
Attention resolutions        & $1/16,1/8$       & Training precision & \texttt{bf16-mixed} \\
Attention head width & 32 & & \\
\hline
\end{tabular}
\end{table}

\subsection{Model Selection}

The configuration in Table~\ref{tab:hparams} was fixed by a two-stage selection, both stages carried out on data disjoint from the test set. Within a training run, the checkpoint is taken at the epoch of minimum validation per-pixel $\chi^2_\nu$ between the prediction and \textit{Euclid}. Across runs, the model size (channel width, multipliers, residual blocks, attention heads) and the noise schedule ($T$, the legacy schedule scale \texttt{beta\_max}, and the number of reverse steps) were tuned over $25$ trials with \textsc{Optuna} \citep{akiba_optuna_2019}. Each trial was evaluated on a held-out $5{,}000$-galaxy validation subset using the S\'ersic index, S\'ersic effective radius, S\'ersic magnitude, and $2\farcs5$-aperture magnitude. For parameter $p$, $b_p$ and $d_p$ denote the median and NMAD of the prediction-minus-\textit{Euclid} residuals. The trial score was
\begin{equation}
S=\sum_p w_p\left[\frac{|b_p|}{s_p}+0.5\max\!\left(0,\frac{d_p}{d_{p,0}}-1\right)\right],
\end{equation}
with weights $(1.5,1.5,1,1)$ in the parameter order listed above; $s_p$ and $d_{p,0}$ are the absolute bias and NMAD fixed from the first completed trial. S\'ersic-derived quantities require clean S\'ersic-fit flags for both the prediction and the \textit{Euclid} reference. The released model is the trial minimizing $S$.

\section{S\'ersic Fitting Configuration}
\label{app:fitting}

Single-component S\'ersic profiles are fitted to the \textit{Euclid} images, E-BGS images, and DESI observations using \textsc{statmorph} \citep{rodriguez-gomez_optical_2019}, with bounds identical across the three datasets. The fit is bounded as $60\le x,y\le68$\,pixel for the center in the native $128\times128$ cutout frame, $0.01\le R_{\rm e}\le64$\,pixel ($6\farcs4$), $0.25\le n\le8$ for the S\'ersic index, and $0.001\le\epsilon\le0.95$ for the ellipticity; position angle, total magnitude, and central amplitude are fitted unbounded. 
Only pixels within the cutout enter the fit; no image values outside the stamp are extrapolated into the likelihood.

\section{Sample Selection Flags}
\label{app:selection}

Training and validation use BGS \emph{targets} selected photometrically from the Legacy Surveys \citep[\texttt{BGS\_BRIGHT}/\texttt{BGS\_FAINT}; $r<20.175$;][]{hahn_desi_2023b}, without a redshift requirement, while the spectroscopic test set is drawn from the DESI DR1 catalog \citep{collaboration_data_2025} as primary galaxy entries with no redshift-fitting failure (\texttt{ZCAT\_PRIMARY = True}, \texttt{SPECTYPE = "GALAXY"}, \texttt{ZWARN == 0}). The corresponding Q1 MER sources are required to satisfy \texttt{VIS\_DET = 1}, \texttt{DET\_QUALITY\_FLAG = 0}, \texttt{SPURIOUS\_FLAG = 0}, \texttt{SPURIOUS\_PROB} $\le0.1$, and \texttt{POINT\_LIKE\_PROB} $\le0.1$.

\section{Multi-Brick Reprojection Details}
\label{app:preproc}

For each \textit{Euclid} MER tile, overlapping DESI bricks are identified from the LIS brick index, and their $r$- and $z$-band images and inverse-variance maps are reprojected from the native $0\farcs262\,\mathrm{pixel}^{-1}$ scale onto the tile grid ($0\farcs1\,\mathrm{pixel}^{-1}$) via world coordinate system (WCS)-based target-to-source coordinates computed with \texttt{scipy.ndimage.map\_coordinates} (\texttt{order=1}). Where a tile pixel is covered by multiple bricks, contributions are combined per-pixel: direct assignment for one brick, mean for two, and median for three or more. The Legacy Surveys coadds used here---DECaLS DR10 and BASS/MzLS DR9---are supplied in nanomaggies per native $0\farcs262$ pixel. The interpolation retains this native-pixel normalization, and integrated fluxes on the $0\farcs1$ grid include the pixel-area factor $(0.1/0.262)^2=0.14568$, equivalently implemented by the photometric zero point 24.59 used for the reprojected DESI measurements. The inverse-variance maps are interpolated with the same coordinate mapping and overlap rule and then converted to RMS maps. These maps provide a diagonal per-pixel uncertainty approximation for source detection and \textsc{statmorph} weighting; covariance introduced by the original coaddition and subsequent reprojection is not explicitly propagated. 
The model-training loss uses a binary validity mask rather than inverse-variance weighting.

\section{Released E-BGS Predictions}
\label{app:release}

\begin{figure*}
    \centering
    \includegraphics[width=0.9\columnwidth]{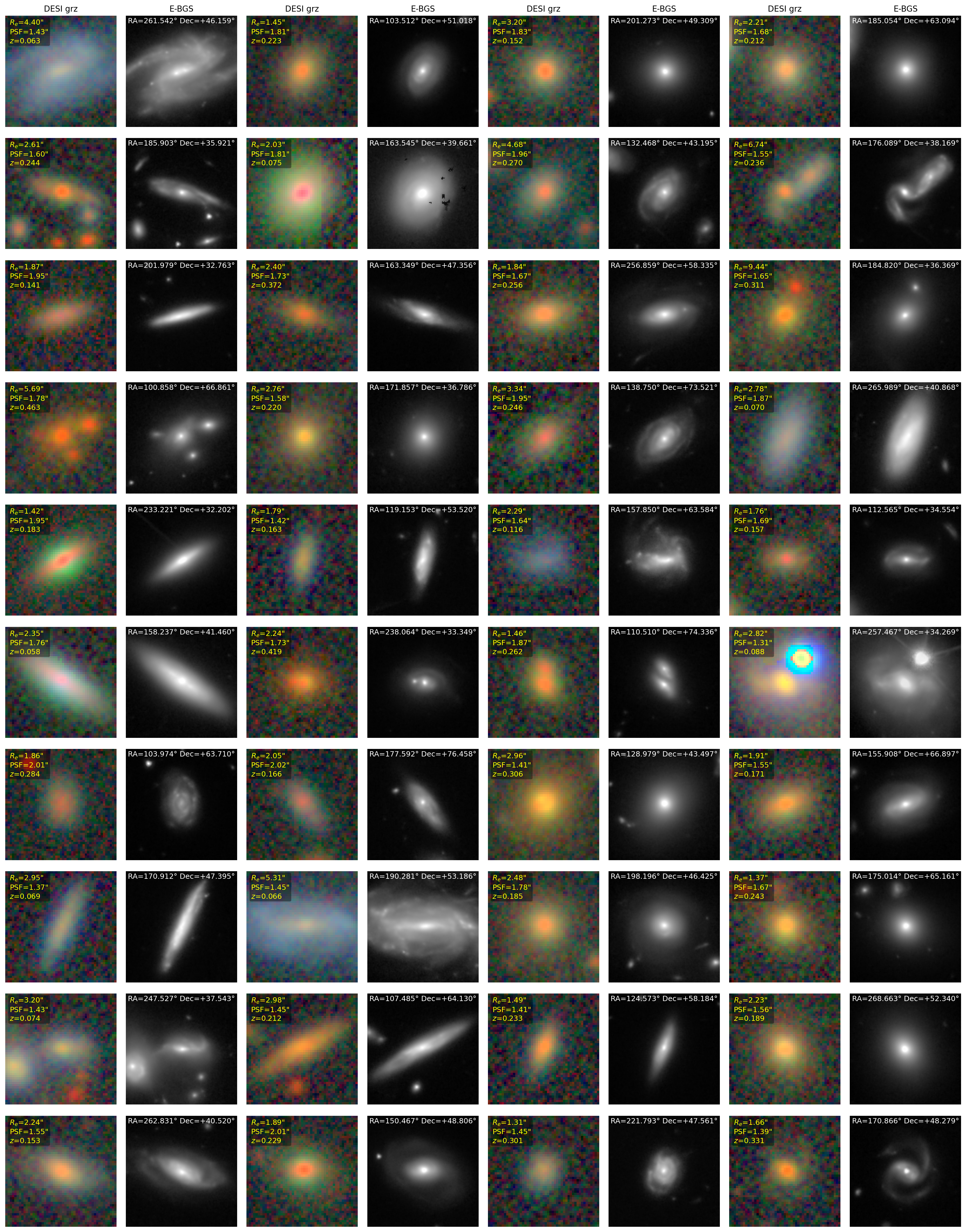}
    \caption{Randomly selected E-BGS predictions over the \textit{Euclid} DR1 footprint, intended} for a future blind comparison with \textit{Euclid} observations. No corresponding public \textit{Euclid} imaging is currently available; these predictions are not substitutes for observed \textit{Euclid} images, and their individual fine-scale features have not been observationally confirmed. Each pair of columns shows the DESI $grz$ composite (used for display only; the model takes the $r$ and $z$ bands) and the corresponding E-BGS prediction (single VIS channel), labeled by R.A.\ and Dec. Annotations on the DESI panels give the single-component S\'ersic effective radius $R_{\rm e}$ measured in the DESI $r$ band, the $r$-band PSF FWHM, and the redshift $z$.
    \label{fig:ebgs_release}
\end{figure*}

\end{document}